# Room temperature coherent control of coupled single spins in solid

T. Gaebel[1], M. Domhan[1], I. Popa[1], C. Wittmann[1], P. Neumann[1], F. Jelezko[1], J.R. Rabeau[2], N. Stavrias[2], A. D. Greentree[2], S. Prawer[2], J. Meijer[3], J. Twamley[4], P. R. Hemmer[5], and J. Wrachtrup[1]

[1] 3. Physikalisches Institut, Universität Stuttgart, Stuttgart, Germany

[2] School of Physics, University of Melbourne, Victoria, Australia

[3] Central laboratory of ion beam and radionuclides, Ruhr University Bochum, Germany

[4] Centre for Quantum Computer Technology, Macquarie University, Sydney, Australia

[5] Department of Electrical and Computer Engineering, Texas A&M University, USA

**Coherent coupling between single quantum objects is at the heart of modern quantum physics. When coupling is strong enough to prevail over decoherence, it can be used for the engineering of correlated quantum states. Especially for solid-state systems, control of quantum correlations has attracted widespread attention because of applications in quantum computing. Such coherent coupling has been demonstrated in a variety of systems at low temperature[1, 2]. Of all quantum systems, spins are potentially the most important, because they offer very long phase memories, sometimes even at room temperature. Although precise control of spins is well established in conventional magnetic resonance[3, 4], existing techniques usually do not allow the readout of *single* spins because of limited sensitivity. In this paper, we explore dipolar magnetic coupling between two single defects in diamond (nitrogen-vacancy and nitrogen) using optical readout of the single nitrogen-vacancy spin states. Long phase memory combined with a defect**

**separation of a few lattice spacings allow us to explore the strong magnetic coupling regime. As the two-defect system was well-isolated from other defects, the long phase memory times of the single spins was not diminished, despite the fact that dipolar interactions are usually seen as undesirable sources of decoherence. A coherent superposition of spin pair quantum states was achieved. The dipolar coupling was used to transfer spin polarisation from a nitrogen-vacancy centre spin to a nitrogen spin, with optical pumping of a nitrogen-vacancy centre leading to efficient initialisation. At the level anticrossing efficient nuclear spin polarisation was achieved. Our results demonstrate an important step towards controlled spin coupling and multi-particle entanglement in the solid state.**

Optically active nitrogen-vacancy (NV) defects in diamond have attracted increasing attention for quantum optical applications following the discovery that the spin states of individual centres can be determined using confocal microscopy[5]. The structure of the NV colour centre is shown in Fig. 1a. It consists of nitrogen and a vacancy in an adjacent lattice site. The NV defect appears in two forms, neutral and negatively charged, and it is negatively charged species that we are concerned here. The ground state is an electron spin triplet ($S=1$)[6], which has been well-characterised in the ensemble limit[7]. The transition between ground and excited spins triplets has a large oscillator strength (0.12), which allows the optical detection of single NV defects. The energy level structure of the NV defect pertinent to our experiments is shown in Fig. 1b. Under optical excitation, the photon scattering efficiency is spin-dependant, allowing readout of the electronic spin state by monitoring the photoluminescence intensity. In addition, fast polarization of the defect via optical pumping is possible[8]. Rabi nutations of single electron and nuclear spins as well as a two-qubit quantum gate have been shown on this system[8,9]. Recently the coupling of the single NV centre electron spin to an electron spin bath of nitrogen impurity spins was demonstrated at room temperature[10]. Room temperature experiments are possible because of the slow spin relaxation.

Different factors contribute to the spin relaxation $T_1$ and dephasing time $T_2$. The interaction with lattice phonons causes population relaxation on a time scale determined by the spin-lattice relaxation time $T_1$. A spin-lattice relaxation time of 1.2 ms at room temperature was reported from NV ensemble measurements[11]. In addition, if other impurity spins are present in the lattice they might form a spin bath and cause the NV spin to decohere via spin-spin interactions. The local field experienced by the NV defect fluctuates because of spin flips among bath spins mediated by dipolar coupling. Such spin flip-flops related to nitrogen spins are confirmed to be the main cause of decoherence for NV centers[12] (single substitutional nitrogen is known to be the major paramagnetic impurity in synthetic diamonds). In order to increase coherence time, the use of higher purity diamond crystals with lower concentration of nitrogen is essential. On the other hand, nitrogen is necessary for formation of NV defects. Auspiciously, NV centres can be created in ultrapure diamond by single ion implantation of nitrogen[13, 14]. This novel technique does not require the presence of nitrogen in the diamond lattice and thus ultrapure (type IIa) diamond can be used as substrate. We can determine the coherence time of such implanted defects. Figure 1c shows a Hahn echo decay time of 350 microseconds from an individual defect. This phase coherence time ($T_2$), is significantly longer than the previously reported ensemble value (50 microseconds for diamonds with very low nitrogen concentration[12]). Presumably, in the present case coupling to $^{13}C$ nuclear spin remains a factor limiting coherence time (natural abundance of $^{13}C$ is 1.1 percent)[15]. The homogeneous line-width of the $^{13}C$ NMR spectrum is 100 Hz at room temperature indicating an average spin flip-flop time of 10 ms[16]. The electron spin feels a change in the local field if a pair of nuclei changes its mutual spin configuration. Flip-flop processes are strongly suppressed in close vicinity to NV centre, because those nuclei experience a strong hyperfine coupling induced energy shift with respect to the spin bath. The decoupling radius $\delta$ is given by[17]

$$\delta = \left[ 2S \frac{\gamma_e}{\gamma_n} \right]^{\frac{1}{4}} a$$

(Here S is the electron spin quantum number, $\gamma_e$ and $\gamma_n$ are the gyromagnetic ratios of electron and nuclear spins, and $a$ is the average nearest neighbour separation between nuclear spins). Substitution of $a$ = 0.44 nm for natural 1.1 % [13]C abundance yields the minimum radius of the frozen core to be 2.2 nm This corresponds to random jumps of the defect centre ESR frequency by about 2.5 kHz, which is in agreement with the experimentally observed phase coherence time. Hence, the availability of isotopically pure diamonds might lead to further increase of $T_2$. In addition, the angular dependence of dipolar coupling can be used for the suppression of decoherence by applying an appropriate magnetic field[18].

The existence of the long coherence times of engineered NV defects allows for a coherent coupling among defects, even in cases of weak coupling strength. Given a $T_2$ of 0.35 ms two electron spins (S=½) should not be separated by more than 15 nm in order for their mutual interaction strength to be larger than the coupling to the bath of [13]C nuclei. Although single NV and N defects can be created one by one using an implantation technique, generation of pairs with intra-pair spacings of only a few nanometres remains challenging. Two factors affect the positioning accuracy. The first is scattering of nitrogen ions in diamond during implantation (straggling). The second factor is ion beam focussing accuracy. Current technology allows implantation of single ions[19] but the spatial implantation accuracy is limited to 20 nm[20]. *Implanting nitrogen molecules* rather than atoms provides a solution to both problems (see Fig. 2a). Although absolute positioning accuracy is still limited by the implanter focus, the *relative* distance (spacing between two defects) is only affected by straggling. Figure 2b shows the distribution of intra-pair spacings for implantations of 14, 10 and 6 keV $N_2$ dimers. The inset shows the fraction of implanted ions with intrapair separations less than 3, 2 and 1.5nm as a function of $N_2$ implantation energy. Note that straggling can be minimized by decreasing the implantation energy resulting in an ever increasing fraction of dimers with intrapair spacings in the regime likely to result in coherent coupling. In the present work, 14 keV $N_2$ dimers were implanted. For this energy the yield of pairs with N-N spacing of 2nm or less is expected to be 1-2% of the implanted $N_2$ dimers

After implanting two closely spaced nitrogen atoms using this molecular implantation techniques the sample was annealed to form NV centres. With the implantation conditions used it was found that about 1% of the molecular implants were converted to a NV-N pair, whilst the rest were left as N-N pair, consistent with the expected yield as described above. Due to the low conversion efficiency of N into NV, no pairs of NV were observed (for details see Methods). As a result, we have concentrated our study on the NV-N pairs. Evidence for NV-N coupling comes from level anticrossing and electron spin resonance experiments. Since the substitutional N defect is an electron spin ½ system, dipolar coupling between the two spins occurs. If the dipolar coupling is weak compared to the NV centre zero-field splitting and Zeeman effect, perturbation theory can be applied to the description of magnetic interactions between the two defects. The Hamiltonian describing the coupled NV-N spin system is[21]

$$H = g_e\beta_e\hat{B}\hat{S}_1 + \hat{S}_1\hat{\hat{D}}\hat{S}_1 + g_e\beta_e\hat{B}\hat{S}_2 + \hat{S}_1\hat{\hat{T}}\hat{S}_2 \tag{1}$$

where $\hat{S}_1, \hat{S}_2$ are spin matrices corresponding to NV and N spins, respectively, $\hat{\hat{D}}$ is the fine structure tensor describing the interaction of the two uncoupled electron spins of the NV defect and $\hat{\hat{T}}$ is the magnetic dipolar interaction tensor. Eigenenergies as a function of external B-field, shown in Fig. 2c, were obtained by diagonalising the spin Hamiltonian. Energy levels are identified according to individual defect spin quantum numbers. According to the energy level scheme, the splitting between doublet components corresponds to dipole-dipole coupling and is expected to be 14 MHz for a defect separation distance of 1.5 nm.

A typical ESR spectrum of a single spin pair is shown in Fig. 3a. As compared to an uncoupled defect, the pairs show a line splitting into two sets of doublets. We find a level anticrossing feature at $B_0$=514G indicating a coupled S=1, S=1/2 system. To demonstrate the coherent nature of the coupling electron spin echo modulation experiments were used. In the spin Hahn echo measurements ($\pi/2 - \tau - \pi - \tau - \pi/2 -$ echo), the amplitude of the echo signal was measured as a function of the pulse separation $\tau$. The $\pi/2$ and $\pi$ pulses in both experiments were 15 and 30 ns, respectively. Therefore the bandwidth is larger than the splitting allowing full excitation of the EPR doublet (AA* transition shown in Figure 3a). The echo envelope shows periodic oscillation (electron spin echo envelope modulation (ESEEM) - Figure 3b). This modulation results from the coherent precession of the NV centre spin in the field created by the nitrogen spin[22]. The related phase acquired by the NV spin is not refocused by the echo sequence resulting in periodic oscillation of the echo amplitude. From a Fourier transformation of the oscillation pattern, the coupling frequency is obtained, yielding an energy splitting between sublevels equivalent to that observed in the EPR spectrum of Figure 3a. The Hahn echo modulation shows not only the frequency which corresponds to NV-N coupling, but also satellites corresponding to the internal hyperfine coupling associated with $^{14}$N. Since any static frequency difference between spins is refocused by the echo sequence, observation of the echo modulation pattern is an unambiguous demonstration of coherent coupling. It is important to note, that no decay of the echo is visible within the measurement time interval.

Only the NV centre couples to the optical field and hence optical initialization can only be applied to the NV centre. A symmetric shape of the ESR doublet in Figure 3a indicates that states $\left|0\right\rangle\left|+\frac{1}{2}\right\rangle$ and $\left|0\right\rangle\left|-\frac{1}{2}\right\rangle$ are populated equally under normal conditions. To polarize (i.e. initialize) the nitrogen spin, resonant spin flip-flop processes induced by dipolar coupling between the NV and N were exploited. Since the spin flip-flop is energy conserving, it is suppressed when the spins are not energetically equivalent, which occurs at low magnetic fields in the NV-N system (see Figure 4a). To achieve polarisation transfer, the frequencies of NV and N spin transitions were tuned into mutual resonance by applying a magnetic field along the symmetry axis of the NV defect. Exact degeneracy is given at the point of level anticrossing B=514 G (see Fig. 2c). The lower graph of Figure 4a shows the evolution of the ESR spectra (transitions AA* in the energy level scheme presented in Figure 2c) as a function of applied magnetic field. When the resonance condition is reached at 514±20 G, the optical pumping on the NV defect rapidly polarises the N defect as well via spin flip-flops (transition between states $\left|0\right\rangle\left|+\frac{1}{2}\right\rangle$ and $\left|-1\right\rangle\left|-\frac{1}{2}\right\rangle$). This polarisation of the N defect is observed by the disappearance of the high frequency component of the ESR spectrum. The time scale for polarisation is given by the coupling strength between NV and N (14 MHz) and the optical pumping rate.

The width of the polarisation transfer resonance is expected to be limited by the homogeneous linewidth of both spin transitions, i.e. some kHz. However, under continuous optical illumination, the resonance of the NV centre broadens because optical pumping disturbs the spin coherence: hence polarisation transfer occurs over a wide magnetic field range because of the overlap between the tails of the resonance lines. To describe the build up of N polarisation we have employed a model including dipolar coupling ($\Delta$), spin-lattice relaxation of nitrogen and NV spins ($\gamma_{SL}^{N}$, $\gamma_{SL}^{NV}$) and optical pumping acting on NV spins ($\gamma_{opt}^{NV}$). The mutual spin flip-flop rate was calculated as product of dipolar coupling ($\Delta = 13$ MHz in the presented case) and the overlap integral $S$ between NV and N lineshapes[23] $S = 1 \left/ \left[1 + \frac{\Delta \cdot D}{2\left(\gamma^{N} \cdot \gamma^{NV}\right)}\right]^{2}\right.$. Here $\gamma^{N}, \gamma^{NV}$ are the dark dephasing rates of N and NV spins, and $D$ is detuning between ESR lines, respectively. The result of calculation without any fitting parameters (the optical polarisation and coherence time of the NV centre was measured independently in pulsed ESR experiments) together with experimental data are shown in Figure 4b.

Detailed examination of ESR spectra reveals not only the disappearance of the high frequency component of ESR doublet, but also an asymmetric narrowing of spectral lines close to NV-N resonance (Figure 4a). This narrowing is related to build up of nuclear polarisation of NV centre. Each ESR line consist of two hyperfine transitions associated [15]N nuclei. Those components are not well resolved in the ODMR spectra presented in Figure 4a because of line broadening associated with optical pumping. Figure 4c show ESR spectra recorded at low optical excitation power. The disappearance of one of the hyperfine transition indicates polarisation of [15]N nuclear spin close to the NV-N resonance. This polarisation of the nuclear spins by an optically aligned electron arise from the "flip/flop" processes involving the simultaneous spin flip of the nuclear and electron spins. This single atom experiment is similar to the nuclear cooling scheme recently demonstrated for quantum dots[24].

The present results demonstrate a first step towards a controlled generation of a set of interacting spins associated with defects in diamond. The system is of potential interest in quantum information processing and communication. Its potential usefulness for solid-state quantum information processing lies in the accessibility of long lived electron spin states by photons. An exquisite control of the spin states has been demonstrated recently[8, 9]. Due to the point-like nature of the defect wave function, the spin coherence is very robust against external perturbations such that coherent control experiments can be carried out at room temperature. Given e.g. the value demonstrated in the present work, the coherence time exceeds the $2\pi$ rotation time by $10^4$. Potentially the present approach is could be used for quantum memory architectures, where the nitrogen defect nuclear spin would be the storage element. In the present work polarization of this nuclear spin via photons has been demonstrated. However, using standard methods of coherence transfer from electron to nuclear spins this could be useful for quantum memory[25, 26] or quantum processing[27, 28].

**Methods.**

To create NV centres in type IIa diamonds (nitrogen concentration <0.1 ppm) we implanted 14 keV $N^+$ ions or $N_2^+$ ions with a dose of ~$2 \times 10^9$ cm$^{-2}$ and annealed the sample for 1 hour in an Ar atmosphere at 900 °C. This procedure lead to formation of NV-N pairs with a yield of 1 percent. For liquid nitrogen substrate implantation temperature the yield is close to 10 %.

Optical detection of single NV-N pairs was performed with 532 nm excitation wavelength using a home-build confocal microscope. To ensure that the optical signal originated from a single NV-N pair, the second order intensity autocorrelation function was measured and the antibunching dip was observed to reach zero at zero delay time.

The hyperfine structure of the ESR spectra was in agreement with implanted N isotope ($^{14}$N or $^{15}$N), indicating that only the implanted nitrogen, and not traces of residual nitrogen present in sample, are responsible for the formation of NV-N pairs. For magnetic resonance measurements the sample was mounted on a miniaturized loop connected to a 40W travelling wave tube amplifier. Results presented in this paper were obtained for $^{14}$NV-$^{14}$N (Figure 3) and $^{15}$NV-$^{15}$N (Figure 4) spin pairs.

**Acknowledgements.** This work was supported by DFG (project "SFB/TR 21"), EU (Integrated Project Qubit Applications (QAP) funded by the IST directorate as Contract Number 015848') and "Landesstiftung B-W" (project "Atomoptik"). The single ion implantation work was supported by by the Australian Research Council, the Australian Government, and the US National Security Agency (NSA), Advanced Research and Development Activity (ARDA), and the Army Research Office (ARO) under contract number DAAD19-01-1-0653 and contract number W911NF-05-1-0284, and DARPA QuIST. We thank G. Tamanyan for technical assistance with the implantations.

**Author information.** The authors declare no competing financial interests. Correspondence and requests for materials should be addressed to F.J. (f.jelezko@physik.uni-stuttgart.de) or J.W. (j.wrachtrup@physik.uni-stuttgart.de).

## Figure caption

**Figure 1| Structure, energy levels and coherence properties of single defects in diamond. a**, Structure of nitrogen-vacancy (NV) centre and subsitutional nitrogen (N) defect centres in diamond. **b**, Energy level diagram of the NV centre. **c,** Hahn echo decay curve of a single NV electron spin recorded at room temperature. During each echo sequence, $\tau_1$ was fixed and $\tau_2$ was varied. Decoherence leads to a reduction of echo amplitude for long time intervals between pulses. The orange curve is an exponential decay fit indicating a phase memory time of 350 microseconds. The grey line shows the fluorescence level corresponding to the full loss of initial polarisation providing a base line for the decay curve.

**Figure 2| Generation of coupled spin pairs. a**, Scheme of molecular implantation leading to formation of NV-N spin pairs. After entering the diamond lattice, the chemical bond of $N_2^+$ molecule is broken and the two N penetrate independently in the diamond. Hence implantation of single $N_2^+$ ions leads to formation of N-N pairs and vacancies (V) in the diamond substrate. Annealing leads to conversion of such pairs into NV-N pairs. **b,** Monte-Carlo simulation of the distribution of intra-pair spacings for implantations of 14, 10 and 6 keV $N_2$ dimers. The inset shows the fraction of implanted pairs with separations between N defects less than 3, 2 and 1.5nm as a function of $N_2$ implantation energy. **c,** Energy levels of dipole-dipole coupled NV-N pair.

**Figure 3| Magnetic resonance on NV-N spin pairs. a**, Optically detected electron spin resonance spectrum of a single NV-N pair in weak magnetic field (40 G). The upper graph shows the ESR spectrum of a single NV centre, indicating two transitions from $m_s$=0 to $m_s$=+/- 1 spin states (A and B). When dipolar coupling is present, each line splits into a doublet (A,A* and B,B* transition shown in the lower graph). **b**, Hahn echo modulation and its Fourier transformation showing coupling between the two spins. Lower graph shows simulation of ESEEM pattern based on dipolar interactions between NV-N electron spins and hyperfine interaction with $^{14}N$ nucleus. Arrow marks coupling frequency between N and NV electron spins. Asterisks indicate frequencies associated with $^{14}N$ hyperfine structure.

**Figure 4| Polarisation transfer between coupled electron spins and build-up of polarization of $^{15}N$ nuclear spin. a**, Evolution of ESR doublet AA* upon varying the external magnetic field. The upper graph represents the scheme of cross polarization of the N electron spin via dipolar coupling to the optically pumped NV centre. **b,** Polarisation P of the N electron spin as a function of

applied magnetic field. Polarisation is defined as $P = \dfrac{I_{A^*} - I_{A^*}}{I_{A^*} + I_A}$ where $I_{A^*}, I_A$ are intensities of ESR spectrum components. The solid red line represents the unfitted model described in text. The upper graph shows the energy level scheme and rates, which were used to describe the build up of N electron spin polarisation. **c.** The curve b shows a high resolution ESR spectra of NV-N pair with low laser excitation power (only transitions A and A* are shown). Each transition shows hyperfine structure (transition 1 and 2 shown in inset) associated with $^{15}$N nuclei. Curve **a** is measured with high laser power. So the lines are broadened and no hyperfine structure is visible. One of the hyperfine components is not visible at energetic resonance between NV and N electron spins (curve c), indicating the build up of nuclear spin polarisation of $^{15}$N of NV centre at NV-N resonance.

Figure 1

**a**

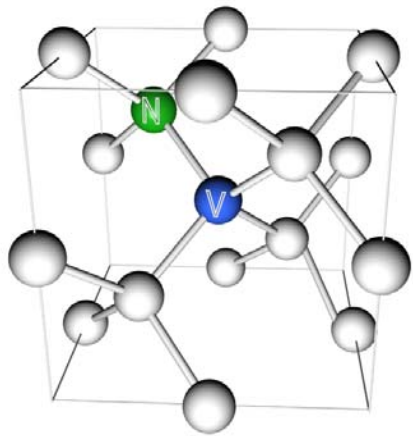

**b**

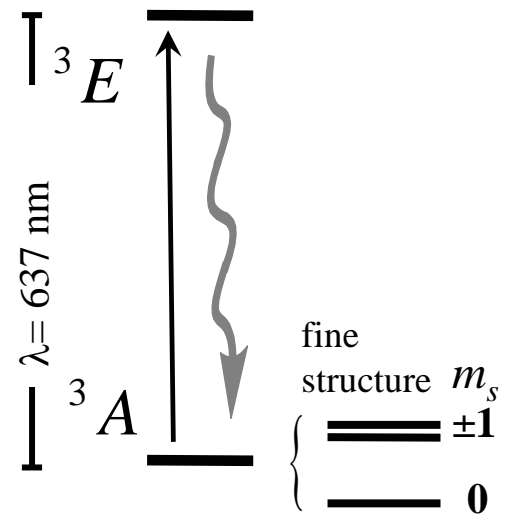

$^3E$

$\lambda = 637$ nm

$^3A$

fine structure $\quad m_s$

$\pm 1$

$0$

**c**

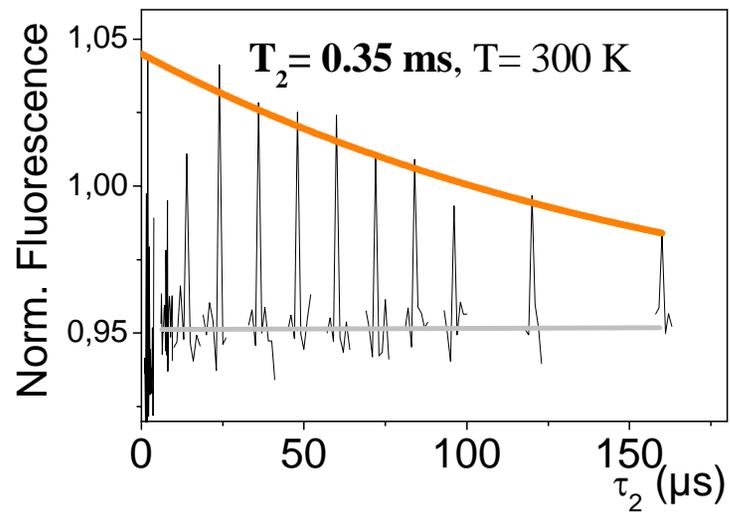

$T_2 = 0.35$ **ms**, T= 300 K

Norm. Fluorescence

1,05

1,00

0,95

0       50      100     150

$\tau_2$ (µs)

Figure 2

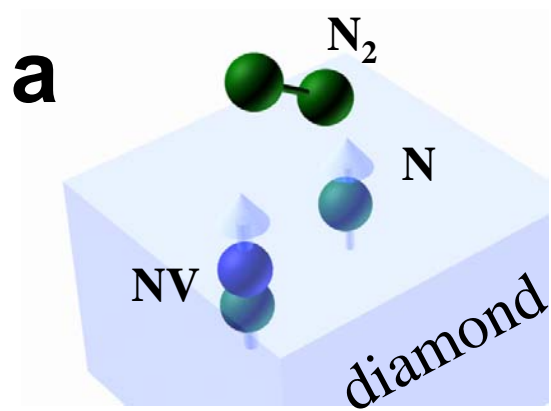

**a**

N₂

N

NV

diamond

**b**

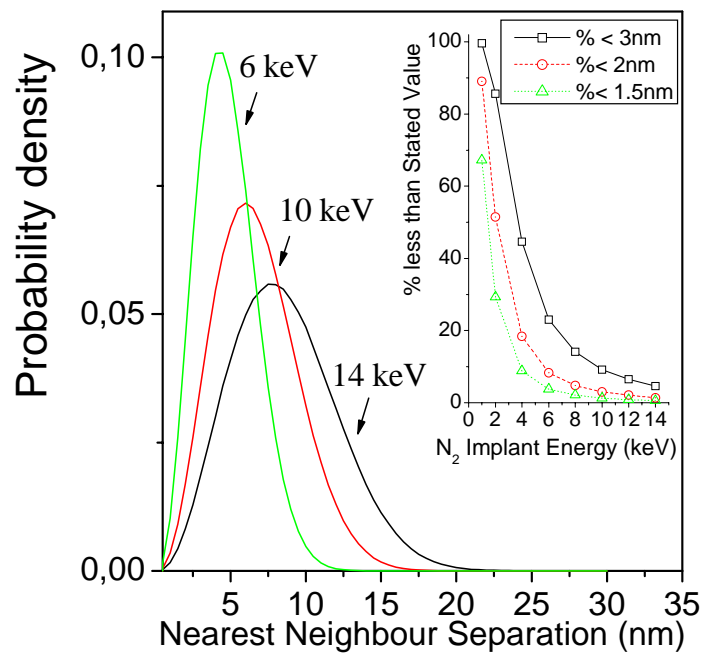

**c**

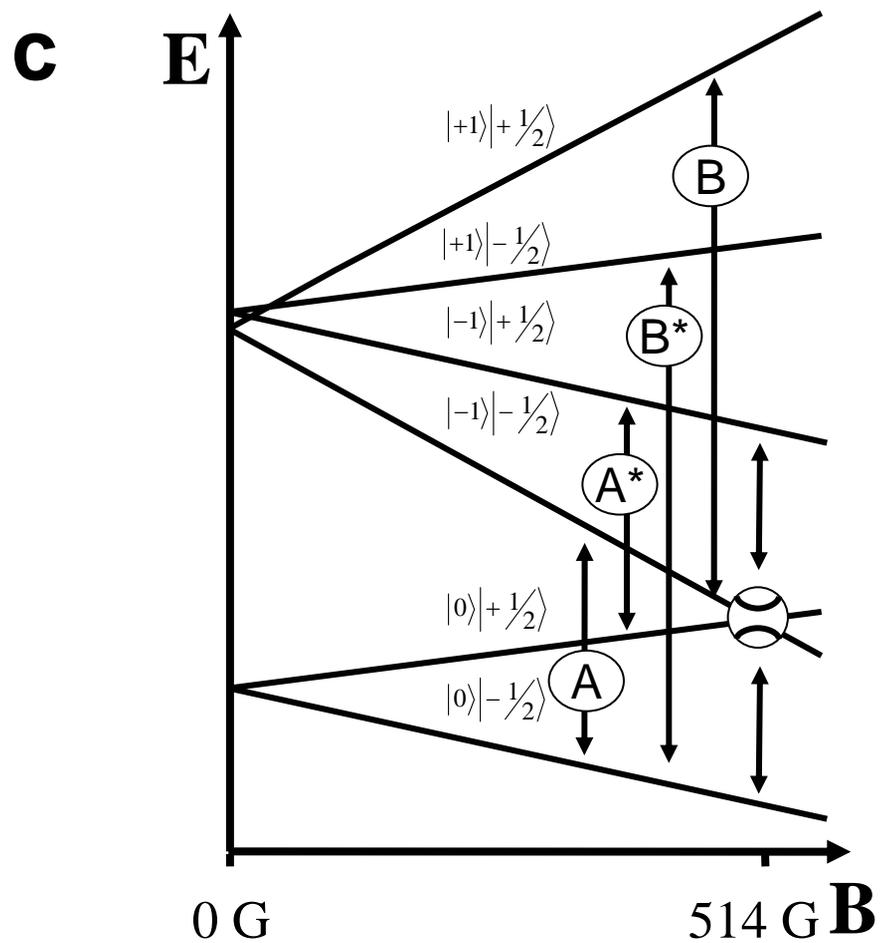

Figure 3

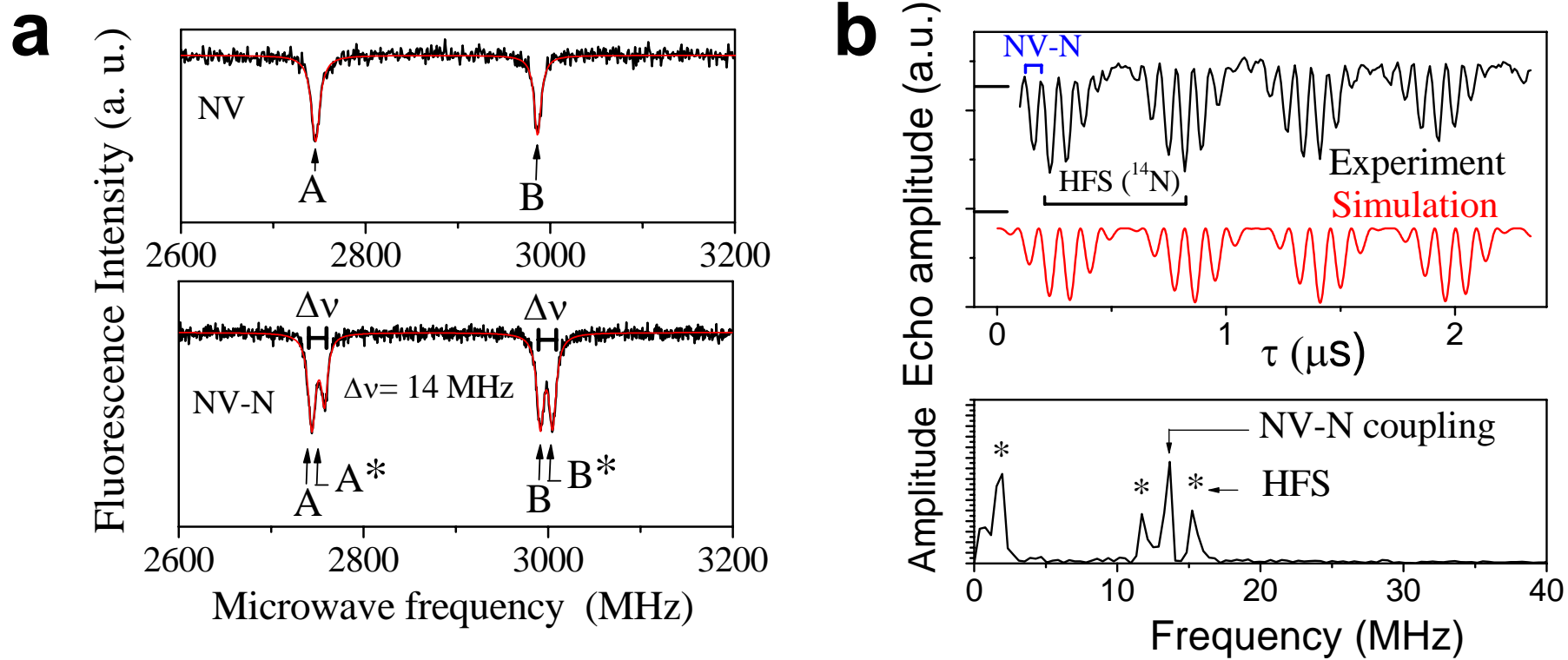

Figure 4

**a**

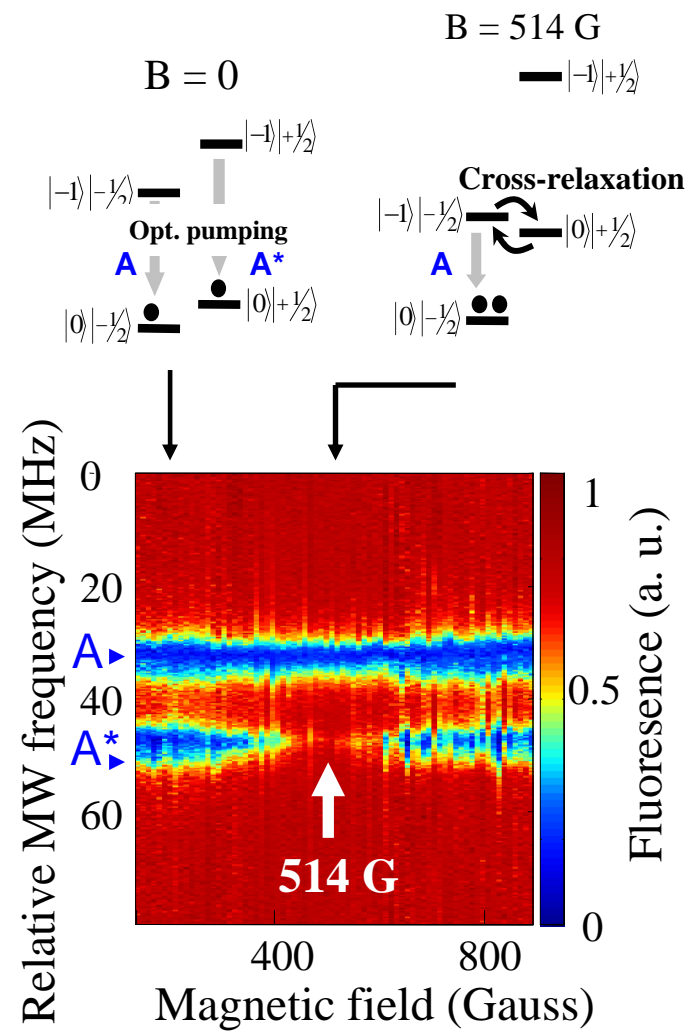

**b**

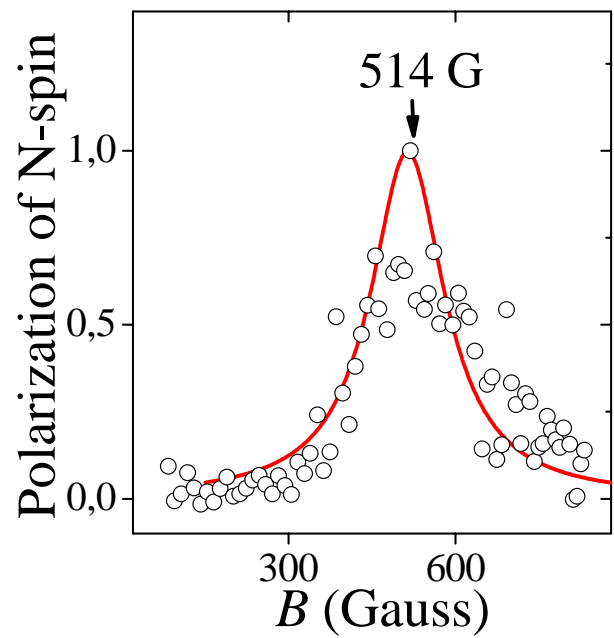

Figure 4

**c**

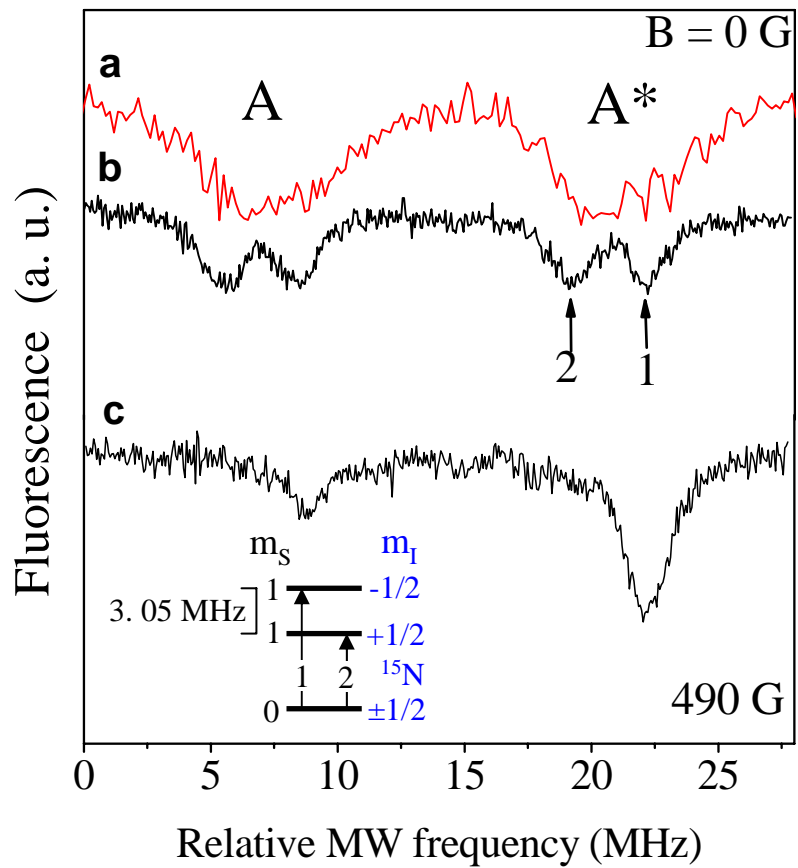